\documentclass[prb,twocolumn,showpacs]{revtex4}

\usepackage{graphicx}
\usepackage{amssymb}
\usepackage{amsmath}

\newcommand{\ud}{\,{\mathrm d}}

\begin{document}

\title{Quasi-uniform in plane magnetization state of thin cylindrical dots 
in a square array and related anisotropy.}

\author{Konstantin L. Metlov}
\affiliation{Institute of Physics ASCR, Prague, Czech Republic 18221; \\ Donetsk Institute of Physics and Technology NAS, Donetsk, Ukraine 83114}
\email{metlov@kinetic.ac.donetsk.ua}
\date{\today}
\begin{abstract}
  The energy (magnetostatic, exchange and Zeeman terms) of a square
  array of cylindrical sub-micron dots made of soft ferromagnetic
  material is calculated analytically and minimized, taking into
  account quasi-uniformity of dots magnetization. The dependence of
  the equilibrium energy of the array on the direction of the
  externally applied magnetic field in the array plane is recovered,
  exhibiting the four-fold anisotropy. The anisotropy constant is
  calculated. Its values for different array geometries are in
  excellent agreement to the recent independent experiments.  New
  eight-fold anisotropy effect is predicted. Theory involves no
  adjustable parameters.
\end{abstract}
\pacs{75.60.Ch, 75.70.Kw, 85.70.Kh}
\keywords{micromagnetics, magnetic nano-dots, quasi-uniform magnetization}
\maketitle

Advances in fabrication of sub-micron ferromagnetic elements and their
arrays produced a wealth of experiments, many of which are still
waiting to be explained by theory. In particular, the first
measurement of the four-fold anisotropy of the spin-wave frequencies
in square arrays of circular permalloy dots was reported in
1997.\cite{MHBBRRDHBCDRCMHH97}

Discovery of the four-fold anisotropy of properties of a square
lattice might seem obvious at first. However, a deeper look reveals
that, while the distribution of stray fields in a square array of
circular magnetic dots depends on orientation of their {\em uniform
magnetization} in the array plane,\cite{M00_dots} the magnetostatic
energy of the array is {\em completely isotropic} for any in-plane
magnetization direction.\cite{G99} It directly follows from Maxwell
equations that interaction of parallel dipoles (which a uniform
magnetization state is) depends only on second powers of direction
cosines and can only produce biaxial anisotropy (uniaxial in the
array plane), even if the array is rectangular. But, as soon as both
periods of the rectangular array are equal, the (in-plane) anisotropy
of the magnetic energy of the uniformly magnetized array vanishes.

Thus, it is clear that the effect must come from some kind of
non-uniform magnetization distribution within the dots. For example,
the two-domain dots (divided by the oppositely-magnetized domains in
half) do show the four-fold anisotropy.\cite{G01_PLA} It can be
doubted, however, that such a configuration would survive the high
(almost saturating) external field of magnetic resonance experiments.

Based on the approximate analytical approach to micromagnetics of thin
flat sub-micron soft ferromagnetic elements,\cite{M01_solitons2} the
ansatz for quasi-uniform magnetization distribution in a circular dot
was proposed recently.\cite{MG04} Using this distribution as a
starting point, the equilibrium (in the Ritz sense) energy of the square
{\em array} of the dots in such a state is calculated below, 
comparing its angular dependence to the experiment.

\begin{figure}[htbp]
  \begin{center}
    \includegraphics[width=7cm]{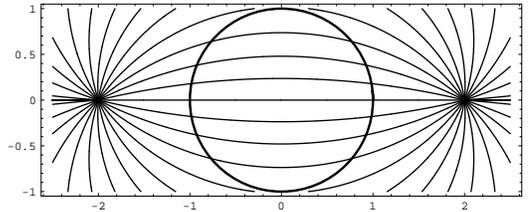}
    \vspace{-0.6cm}
  \end{center}
  \caption{The quasi-uniform magnetization
    distribution (\ref{eq:wzarr}) with $p=2$. Lines
    inside the particle (its contour is outlined by the circle) are
    tangential to the magnetization vectors.}
  \label{fig:illustr}
\end{figure}

The magnetization distribution\cite{M01_solitons2} can be
expressed in terms of the complex function of complex variable:
\begin{equation}
  \label{eq:wzarr}
    w(z,\overline{z}) = e^{\imath \varphi_0} 
    \sqrt{\frac{p^2-e^{- 2\imath\varphi_0} z^2}
      {p^2-e^{2\imath\varphi_0}\overline{z}^2}},
\end{equation}
where $z=X/R+\imath Y/R$, $X$, $Y$ are Cartesian coordinates on the
dot's face, $R$ is the dot radius, $\imath=\sqrt{-1}$, line over a
variable denotes the complex conjugation.
Components of magnetization unit vector $\vec{m}$ are expressed
through $w(z,\overline{z})$ as $m_X+\imath m_Y = 2 w / (1 + w
\overline{w})$ and $m_Z=(1 - w \overline{w})/(1 + w \overline{w})=0$.
The dimensionless parameter $p>1$ describes displacement of skyrmions
from the cylinder's side, their centers (zeros of $w(z,\overline{z})$)
are located at $X =\pm p R$, $Y=0$. The value of $p$ is to be found
from minimization of the total energy of the array.
For $p\rightarrow\infty$ the magnetization distribution is uniform,
for finite $p$ the corresponding ``leaf'' quasi-uniform magnetization
state is shown in Fig.~\ref{fig:illustr}.
Compared to the original ansatz in
Ref.~\onlinecite{MG04} the expression (\ref{eq:wzarr}) 
allows the average magnetization of the dot to point in arbitrary
direction at an angle $\varphi_0$ w.r.t one of the array axis (this
was unnecessary complication for a single dot\cite{MG04}).

To evaluate the magnetostatic energy of the array of such dots
the magnetic charges formalism is used. It is based on the assumption
that the problem is static and there are no currents in the system.
This allows to express the stray magnetic field through a scalar
potential, created by a system of ``fictious'' (introduced for
mathematical convenience) magnetic charges. The volume density of
these charges is $\sigma_V= - \mathrm{div} (\vec{m})$. On the boundary
of magnetic material $\sigma_V$ reduces to the surface charge with the
density $\sigma_S= (\vec{m}\cdot\vec{n})$ proportional to the
(boundary) normal component of the magnetization vector. Both
densities are normalized here by the saturation magnetization of the
dot's material ($M_s$). Inside of a dot with magnetization distribution
(\ref{eq:wzarr}) they are:
\begin{eqnarray}
  \label{eq:volume}
  \sigma_V & = & \frac{2 r \cos(\varphi - \varphi_0)}
  {\sqrt{r^4 + p^4 - 2 p^2 r^2 \cos 2 (\varphi - \varphi_0)}}, \\
  \label{eq:surface}
  \sigma_S & = & \frac{(p^2 - 1) \cos(\varphi - \varphi_0)}
  {\sqrt{1 + p^4  - 2 p^2 \cos 2 (\varphi - \varphi_0)}},
\end{eqnarray}
where $r \leq 1$ and $0<\varphi \leq 2\pi$ are the polar coordinates
on the dot's circular face, normalized by the dot radius. These are
the rotated charge densities of Ref.~\onlinecite{MG04}.

To fully take into account periodicity of the array and all the
implied interactions between dots, the density of magnetic charges is
represented by its Fourier components:
\begin{eqnarray}
  \!\!\!\!\widetilde{A}_{n,m}(k_z)\!\! 
  & = &\!\!\frac{\rho\,\widetilde{a}_{n,m}}{T}
   \!\!\!\!\!\!\!\int\limits_{-L_Z/2}^{L_Z/2}\!\!\!\!\!\!\!e^{-2\pi\imath k_Z Z} \!\ud Z
   \! =\!
   \frac{\widetilde{a}_{n,m}\rho \lambda \sin k_Z\!L_Z\!\pi}{k_Z L_Z \pi}\\
   \label{eq:fourier2d}
  \!\!\!\!\widetilde{a}_{n,m}\!\! & = &  
  \!\!\int\limits_0^{2\pi}\ud \varphi 
  \left(
    \sigma_S\,F|_{r=1} + 
    \int\limits_0^1 r \ud r\,\sigma_V\,F 
  \right),\\
  \label{eq:fourierbasis}
  \!\!\!\!F\!\! & = & \!\! e^{-2\pi\imath \rho(n X + m Y)} 
  = e^{-2\pi\imath\rho r(n \cos\varphi + m\sin\varphi)},
\end{eqnarray}
where $\rho=R/T$, $\lambda=L_Z/T$, $T>2R$ and $L_Z$ are the array
period and thickness, $\widetilde{a}_{0,0}\equiv 0$ by construction.

Because the Fourier basis functions $F$ are also the eigenfunctions of
the Laplace differential operator, the Poisson equation (resulting
from the Maxwell equations under assumptions of the magnetic charges
formalism) for the scalar potential becomes algebraic for its Fourier
harmonics and can be readily solved. Normalized magnetostatic energy
per array cell volume due to the interaction of the stray field with
the magnetic charges $e_M=E_M/(\mu_0\gamma_B M_S^2 T^2 L_Z)$ is then
\begin{eqnarray}
  \label{eq:emstraight}
  e_M\! = \!\frac{\rho^2\lambda^2}{8\pi^2}
  \!\!\!\sum\limits_{n=-\infty}^{\infty}\sum\limits_{m=-\infty}^{\infty}
  \int\limits_{-\infty}^{\infty}\!\!\! \ud \xi 
  \frac{|\widetilde{a}_{n,m}|^2}{\lambda^2(n^2\!+\!m^2)\!+\!\xi^2}
  \frac{\sin^2 \xi \pi}{(\xi\pi)^2},
\end{eqnarray}
where
$|\widetilde{a}_{n,m}|^2=\widetilde{a}_{n,m}\widetilde{a}_{-n,-m}$,
$\xi=k_Z L_Z$, $\mu_0$ is permeability of vacuum in SI units,
$\gamma_B$ is a units-dependent factor\cite{Aharoni_book} equal to $1$
in SI (in CGS $\mu_0=1$, $\gamma_B=4\pi$). The integral can be taken,
allowing to represent
\begin{eqnarray}
  \label{eq:emnokz}
  e_M & = & \frac{\rho^2}{8\pi^2}
  \!\!\!\sum\limits_{n=-\infty}^{\infty}\sum\limits_{m=-\infty}^{\infty}
  \!\!\!\!\!
  \frac{|\widetilde{a}_{n,m}|^2 \mathrm{f}(2\pi\lambda \sqrt{n^2\!+\!m^2})}
  {n^2\!+\!m^2}, \\
  \mathrm{f}(x) & = & 1+\frac{e^{-x}-1}{x}.
\end{eqnarray}

To evaluate the 2-d Fourier components (\ref{eq:fourier2d}) let us
first note that the denominator of both surface and volume charges can be
expanded into the power series in $1/p^2$ (a small parameter
in case the magnetization state in the dot is quasi-uniform) using the
identity for the generating function of the Legendre polynomials $P_i(x)$
\begin{equation}
  \label{eq:legendregen}
  \frac{1}{\sqrt{r^4+p^4 - 2 p^2 r^2 \cos \psi}}
  =
  \frac{1}{p^2}
  \sum\limits_{i=0}^\infty 
  \frac{r^{2 i}P_i(\cos \psi)}{p^{2 i}}
\end{equation}
where $\psi=2(\varphi\!-\!\varphi_0)$. Using the identities for the
generating functions of Bessel's functions (formulas 6.521 in
Ref.~\onlinecite{Gradshtejn_Ryzhik}), the Fourier basis
(\ref{eq:fourierbasis}) can be represented as
\begin{equation}
  \label{eq:besselgen}
  F\!\! = \!\!\sum\limits_{k=-\infty}^{\infty}\sum\limits_{l=-\infty}^{\infty}
  \!(-\imath)^k (-1)^l J_k(\alpha\, n) J_l(\alpha\, m)
  e^{\imath (k+l) \varphi},
\end{equation}
where $\alpha=2\pi\imath\rho r$ and $J_k(x)$ are the Bessel's functions of the first kind.

In such a representation it is easy to evaluate the angular integral
(\ref{eq:fourier2d}), which, for a given $i$, is non-zero only
for $k+l=\pm(2j+1)$ with $j=1,2,\ldots,i$. This allows to remove (by
making it finite) one infinite summation from a triple $i$, $k$, $l$
sum, resulting from expansions (\ref{eq:legendregen}),
(\ref{eq:besselgen}). Then, using Bessel's summation theorem, each
of the remaining infinite sums over $k+l=(2j+1)$ and $k+l=-(2j+1)$ can
be evaluated and combined, yielding the representation of
$\widetilde{a}_{n,m}$ term for a given $i$ in terms of finite sum of
Bessel's functions of the orders $2j+1$ for $j=1,2,3,\ldots,i$. The
resulting terms can be easily integrated in $r$ for the part of
the integral (\ref{eq:fourier2d}), representing the volume charges.
This calculation is quite voluminous in general.\cite{packageFourier}
Fortunately, to obtain the expansion of the magnetostatic energy up to
the $1/p^4$ (the first neglected term is $~\sim 1/p^6$) it is
sufficient to evaluate the volume charges Fourier components for
$i=0,1$ and the surface charges for $i=0,1,2$ (because of $p^2$
in Eq.~\ref{eq:surface}).

The final result for the magnetostatic energy can be represented as
\begin{eqnarray}
  \label{eq:msfinal}
  \!\!\!\!e_M\!\! & = & \!\!e_0 \!+\!
  \frac{e_1\!+\!\cos(4 \varphi_0) e_{1a}}{p^2} \!+\!
  \frac{e_2\!+\!\cos(4 \varphi_0) e_{2a}}{p^4} \!+\! \ldots,\\
  \!\!\!\!e_i\!\! & = & \!\!\rho^2 \!\!
  \sum\limits_{m=0}^{\infty}\sum\limits_{n=1}^{\infty}\!
  \frac{K_i(\rho k_{m,n},\psi_{m,n})}{m^2+n^2} f(\lambda k_{m,n}),
\end{eqnarray}
where $k_{m,n} = 2\pi\sqrt{m^2+n^2}$, $\psi_{m,n} = \frac{1}{2}\arctan(\frac{2 m n}{n^2-m^2})$,
\begin{eqnarray}
  K_0(x) & = & J_1^2(x), \\
  K_1(x) & = & J_1(x) J_3(x), \\
  K_{1a}(x,\psi) & = & J_1(x) J_3(x)\cos(4\psi), \\
  K_2(x) & = & - 2 J_0(x) J_3(x)/x,\\
  K_{2a}(x,\psi) & = & (J_3(x)^2 +2 J_1(x)J_5(x))\cos(4\psi)/2.
\end{eqnarray}
Functions $e_i=e_i(\rho,\lambda)$ depend only on geometry of the
problem and take into account the self-energy of the volume and
surface charges, as well as interaction between them, both in a single
dot and across all the dots in the array.

These expressions can be verified by renormalizing them to the unit of
magnetic dot volume (as opposed to the volume of lattice cell) and
taking the limit $T\rightarrow\infty$. In this case $e_{1a}=e_{2a}=0$,
and $e_0$, $e_1$ and $e_2$ become identical to the corresponding
functions $e^\parallel$, $e_2$ and $e_4$ of Ref.~\onlinecite{MG04}.
Because the same thoroughly verified automatic\cite{packageFourier}
procedure was used in deriving all the functions $K_i$, including the
angular ones (marked with ``a''), it can be expected with a
high degree of confidence that all the expressions for
$e_i(\rho,\lambda)$ are correct.

Looking at (\ref{eq:msfinal}) it is immediately seen that for
uniformly magnetized dots (when $p\rightarrow\infty$) the
magnetostatic energy shows no angular dependence. The function $e_0$
is, thus, the energy of a square lattice of uniformly magnetized dots,
which can be independently verified.

Compared to the above, evaluation of the exchange and Zeeman energy
terms is trivial. Both contain no interaction between dots and can be
evaluated for each dot separately. Actually, both were already
calculated in Ref.~\onlinecite{MG04}. Renormalizing (2) from
Ref.~\onlinecite{MG04} as $e_E=E_E/(\mu_0\gamma_B M_S^2 T^2 L_Z)$ and
using the expression for the average magnetization (16) therein we get
up to $1/p^4$:
\begin{equation}
  \label{eq:exchzeeman}
  e_E \!+\! e_Z \!=\! \frac{\pi}{\gamma_B \tau^2 p^4} \!-\! 
  \pi \rho^2 h \cos (\varphi_0\!-\!\varphi_1)
  \left(\!\!
    1 - \frac{1}{12 p^4}
  \!\!\right) \!+\! O\left(\!\frac{1}{p^8}\!\!\right)
\end{equation}
where $\tau=T/L_E$, $L_E=\sqrt{C/(\mu_0 M_S^2)}$ is the exchange
length of dot's material, $h=H/(\mu_0\gamma_B M_S)$ is the normalized
field magnitude, and $\varphi_1$ is the angle of the in-plane applied
field with respect to the lattice.

Now it remains to minimize the sum of (\ref{eq:msfinal}) and
(\ref{eq:exchzeeman}) to find the equilibrium values of $p$ and
$\varphi_0$. This results in two solutions for $\varphi_0$ separated
by a small angle from the direction of the applied field $\varphi_1$,
meaning that there will be hysteresis, depending on whether the
current direction of the applied field was approached from bigger or
smaller angles. Because the typical magnetic resonance measurement is
performed in high sub-saturating magnetic field, this hysteresis can
be expected to be small. Let us neglect it here for simplicity by
putting $\varphi_0=\varphi_1$. Minimizing we get
\begin{equation}
  \label{eq:peq}
  p^2 = \frac{h \pi \rho^2 + 12 (\pi/(\gamma_B \tau^2) + e_2 + e_{2a} \cos 4\varphi_1)}
  {6(-e_1 - e_{1a} \cos 4\varphi_1)}.
\end{equation}
Let us assume for now that the denominator here is always positive
(that is $e_1<0$, $e_{1a}>0$ and $|e_{1a}|<|e_1|$), which is the case
for all the experiments analyzed below. Breaking of this assumption
will be discussed at the end. The value of $p$ completely defines
magnetic structure of dots in the array.  Substituting it back, the
equilibrium energy can also be expressed straightforwardly, but let us
now concentrate on its angular dependence.

The largest (besides the constant term) Fourier harmonic in the
angular dependence of the equilibrium energy is proportional to $\cos
4\varphi_1$, which can be clearly classified as the four-fold
anisotropy. The higher harmonics, while modifying slightly the
dependence itself, do not alter positions of extrema on the angular
dependence with minimums located at $\varphi_1=\pi/4$, $3\pi/4$,
$5\pi/4$, $7\pi/4$ and maximums at $\varphi_1=0$, $\pi/2$, $\pi$, and
$3\pi/2$. Let us, thus, introduce the fourfold anisotropy constant,
representing the equilibrium energy of a single dot (that is,
normalized by the dot volume, which is the volume of magnetic material
in the array cell) as $E=E_0 + K_4 \cos 4\varphi_1$ with $K_4>0$.  The
value of $K_4$ is the half of the equilibrium total energy
(renormalized by the dot volume) difference between configurations at
$\varphi_1=0$ and $\varphi_1=\pi/4$.  Further, noting that in the
resonance experiments the high external field plays a more notable
role in stabilization of the ``leaf'' state than the exchange
interaction (stabilizing it in the absence of the applied field), a
limit of $\tau\rightarrow\infty$ is taken to simplify the expression
into
\begin{equation}
  \label{eq:k4notau}
  K_4 \!=\!\frac{\mu_0\gamma_B M_s^2}{2\pi\rho^2}\!\!\left(\!\!
    \frac{3(e_1\!-\!e_{1a})^2}{12(e_2\!-\!e_{2a})\!+\!\pi h \rho^2}\! - \!
    \frac{3(e_1\!+\!e_{1a})^2}{12(e_2\!+\!e_{2a})\!+\!\pi h \rho^2}
  \!\!\right)
\end{equation}
The full expression, more accurate for the case of the array period
comparable to the exchange length, can be easily obtained by the
reader (it is also plotted in figures here). Nevertheless, for the
subsequent comparison to experiments the precision of
(\ref{eq:k4notau}) is sufficient. At large applied fields
$H/(\mu_0\gamma_B M_S) \gg 1$ it is possible to expand $K_4$
asymptotically as
\begin{equation}
  \label{eq:k4asympt}
  K_4 = - \mu_0 \gamma_B M_S^2 \frac{6 e_1 e_{1a}}{\pi^2\rho^4 h}+O\left(\frac{1}{h^2}\right).
\end{equation}

\begin{figure}[t]
  \begin{center}
    \includegraphics[width=8cm]{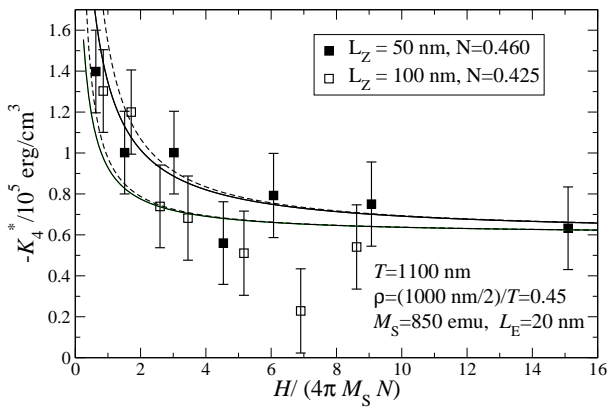}
    \vspace{-0.6cm}
  \end{center}
  \caption{Dependence of the four-fold anisotropy constant $K_4^{*}$
    on the magnitude of the in-plane magnetic field. Points are from
    Ref.~\onlinecite{MHBBRRDHBCDRCMHH97}. The solid lines
    (\ref{eq:k4notau}) and the full expression (without taking
    $\tau\rightarrow\infty$ limit) are indistinguishable. Dashed lines
    are from Eq.~\ref{eq:k4asympt}. The upper set of lines corresponds
    to $L_Z=100\mathrm{nm}$, the other to $L_Z=50\mathrm{nm}$. The
    values of the in-plane demagnetizing factor $N$ are shown in the
    legend.}
  \label{fig:MHBBRRDHBCDRCMHH97}
\end{figure}

Let us now apply the obtained knowledge to describe experiments found
in literature. The first and foremost is the experiment found in
Ref.~\onlinecite{MHBBRRDHBCDRCMHH97}. To interpret it, the authors
added the anisotropy of the form $K_{4}^{*}\sin^2\psi\cos^2\psi$
($\psi$ is the angle between the magnetic moment in the dot and
the lattice axis) into their numerical program for finding the
resonance frequencies and fitted the angular dependence of the spectra
to determine the $K_4^{*}$. Because the magnetization of the dot is
close to uniform ($\psi=\varphi_1$) this term produces the shift in
the total dot's energy, followed by the measured spin-wave mode.
Due to $\sin^2\varphi_1\cos^2\varphi_1$ oscillating between $0$ and
$1/4$, while $\cos 4\varphi_1$ between $-1$ and $1$ with maximums of
one corresponding to minimums of another (and vice versa), there is
additional factor $-8$ linking the anisotropy in
Ref.~\onlinecite{MHBBRRDHBCDRCMHH97} to the definition here.
Another peculiarity is that the value of $K_4^{*}$ in
Ref.~\onlinecite{MHBBRRDHBCDRCMHH97} tends to a constant value at
large applied fields. This can not be explained by any model of the
quasi-uniform state of infinite array, which at $h\rightarrow\infty$
must transform into the uniform state, showing no anisotropy. Thus, it
seems another factor (probably due to the array shape effect) is
present in the experiment. Assuming this factor is
field-independent, let us simply add it, expressing $K_4^{*}=-8 K_4 -
0.6\cdot 10^{5} \mathrm{erg}/\mathrm{cm}^3$, plotted in
Fig.~\ref{fig:MHBBRRDHBCDRCMHH97}
(taking $M_s=850$ emu, typical for permalloy, but not specified in
Ref.~\onlinecite{MHBBRRDHBCDRCMHH97}) along with the experimental
data. The agreement seems to be rather good, noting that the precision
of the experiment did not allow\cite{MHBBRRDHBCDRCMHH97} to reliably
resolve between dependencies of the $K_4^{*}$ for two considered
arrays.

\begin{figure}[t]
  \begin{center}
    \includegraphics[width=8cm]{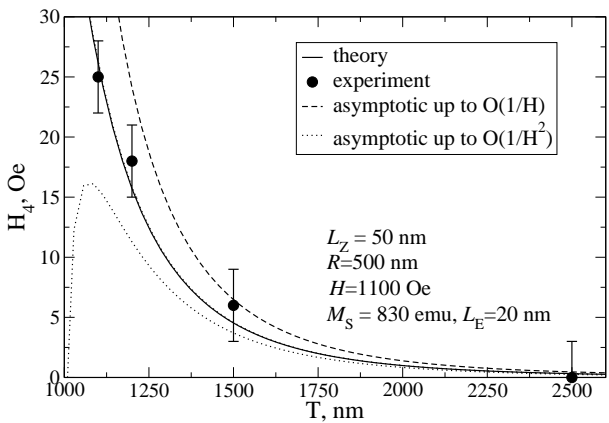}
    \vspace{-0.6cm}
  \end{center}
  \caption{Dependence of the four-fold anisotropy field $H_4$ on the
    inter-dot spacing. Points are from Ref.~\onlinecite{Gleb06}.  The
    solid line plots the Eq.~\ref{eq:k4notau} (again, the finite
    $\tau$ case is indistinguishable). Dashed and dotted lines show
    the asymptotes of (\ref{eq:k4notau}) up to $O(1/h)$
    (Eq.~\ref{eq:k4asympt}) and $O(1/h^2)$ respectively. The value of
    $L_E$ obtained from the measured\cite{Gleb04} exchange constant ($A=C/2$).}
  \label{fig:Kakazei}
\end{figure}

A much more precise recent experiment was done using FMR
technique.\cite{Gleb06} In FMR it is impossible to change the value of
the applied field, which was fixed at $1100\mathrm{Oe}$, but the
authors tracked the anisotropy field as a function of the array
geometry, measuring different arrays with periods from $1100$ to
$2500$ nm. The anisotropy field can be expressed through the
anisotropy constant as $H_4 = 2 K_4/M_S$, which is plotted in
Fig.~\ref{fig:Kakazei}
along with the array parameters and the
experimental data from Ref.~\onlinecite{Gleb06}. The agreement is
excellent for the case of $K_4$ calculated from (\ref{eq:k4notau}),
whereas the simplified asymptotic expression (\ref{eq:k4asympt}) and
its higher order equivalent are clearly not good enough at such a
small field.

Let us now look back at (\ref{eq:peq}). It turns out that it is quite
possible for the denominator to become zero and negative. This happens
for dot geometries approximately defined by the numerical fit
$R_{cr}/T = 0.51 - 0.37 \cdot (L_z/T) + 0.13 \cdot (L_z/T)^2$ in
$0<L_z/T<1$ with fixing the resulting $R_{cr}/T$ to 0.5 when it
exceeds this value. At radii above $R_{cr}$ the parameter $p$ becomes
imaginary for some directions of the applied field, which corresponds
to the transition to the so-called ``flower'' state. In this state (as
opposed to initially considered ``leaf'') the magnetization vectors
diverge {\em outwards} of the line through the dot center parallel to
the applied field. All the expressions here still apply in the case of
``flower'' (with imaginary $p$), but, because of the additional
``flower''$\leftrightarrow$``leaf'' transition the dots array with the
$R>R_{cr}$ shows the additional eight-fold (!) anisotropy of the form
$K_4 \cos4\varphi_1 + K_8 \cos 8 \varphi_1$ with $K_4$ still given by
(\ref{eq:k4notau}) and $K_8<0$. This transition and the resulting
anisotropy will be explored in the forthcoming extended paper on the
subject.

To conclude, the considered problem belongs to the class, where the
leading order energy contribution vanishes due to symmetry, and the
system is governed by higher order energy terms. In most other cases
details of the quasi-uniform magnetization distribution make only
small (and often negligible) corrections to the leading order results,
but here they are responsible for the completely new effect, absent in
the leading order. Apart from quantitatively describing the four-fold
anisotropy in the magnetic dot arrays, the presented theory asks for
experiments on thicker dots, which should reveal the new eight-fold
anisotropy effect. This theory can be expected to describe it
quantitatively.

Support by the internal project K1010104 is appreciated. I would like
to thank Vladimir Kambersky for valuable remarks, especially for
doubting the initial conjecture that $-e_1\pm e_{1a}>0$, which pushed me
for a recheck and the discovery of the ``flower'' state in
the system.


\begin{thebibliography}{11}
\bibitem[{pac()}]{packageFourier}
\bibinfo{note}{I have written a Mathematica package to do it for a given $i$,
  which I can share upon request}.
\end{thebibliography}

\end{document}